\documentclass[aps,preprint]{revtex4}
\usepackage{epsf}
\newcommand{\tr}{\mbox{Tr} }
\newcommand{\ket}[1]{\left | #1 \right \rangle}
\newcommand{\bra}[1]{\left \langle #1 \right |}
        
\begin{document}

\title{Alternative Method for Quantum Feedback-Control}

\author{Julian Juhi-Lian Ting}
\email{jlting@chez.com}
\homepage{http://www.chez.com/jlting}
\affiliation{Nanomaterials Laboratory, National Institute for Materials Science, \
3-13 Sakura, Tsukuba, Ibaraki 305-0003, Japan}

\date{\today}

\begin{abstract}
A new method for doing
feedback control of single quantum systems 
was proposed. Instead of
feeding back precisely the process output,
a cloning machine served to obtain the feedback signal and the output.
A simple example was given to demonstrate the method.
\end{abstract}

\maketitle

\section{Introduction}
It is a dream for
both theorists and experimentalists to be able
to control single quantum systems, in particular quantum gates
for quantum computing.
Uchiyama described how to achieve open-loop control for spin-boson model in the first day of the symposium\cite{U}.
Among interesting works on open loop controls,
Viola and Lloyd introduced quantum bang-bang control\cite{VL},
whereas Mancini {\it et al.} considered stochastic resonance to control quantum systems\cite{MVBT}.
However,
as in classical control theories,
open-loop  control is of limited use
because in most cases
it is merely a blind control.

\begin{figure}
\epsfxsize=7.5cm\epsfbox{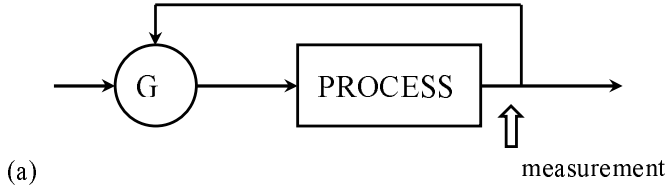}
\vskip 0.5cm
\epsfxsize=7.5cm\epsfbox{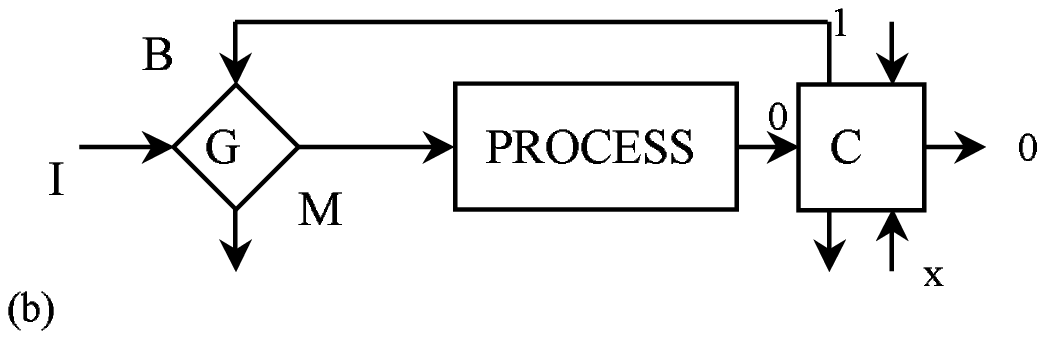}
\vskip 0.3cm
\caption{Comparison of standard (classical or coherent) feedback 
control scheme, (a), and
the proposed control scheme, (b).
Instead of feedback the precise process
output, we feed back the cloned output and output
a cloned process state. 
The cloner, $C$,  has three pairs of input and output of which
"0" is the input state, "1" is the cloned state, and "x" is the 
ancilla (the cloning machine).
For simplicity, actuators are taken to be
an unity function, which made no change to the feedback state,
without lost of generality.}
\label{fig1}
\end{figure}

Feedback control is not only experimentally difficult, but even theoretically
impossible, because a measurement at the output collapses the system's
coherence\cite{VN}.
Wiseman made many attempts to achieve feedback control, based mainly on
the homodyne method of detection\cite{WMW, WM, WWM}.
Doherty {\it et al.} used state estimation methods\cite{DHJMT}.
These methods
with some measurements 
at the output stage to have feedback signals can be summarized
as shown in Fig.(\ref{fig1}a).
However, are such measurements necessary?
Recent works on quantum information theories indicate that 
having signals need not destroy coherence.

In this paper the idea of feeding back precisely the process output
is renounced.
Instead, a quantum cloning machine is placed at the
output side, as shown in Fig.(\ref{fig1}b).
According to  quantum-cloning theorems, it is possible to make either
imperfect copies with unity probability\cite{BH,C}
or perfect copies with probability less than unity\cite{DG}.
In the later case, a measurement is necessary, whereas
the former
cloning machine,
which  can be decomposed into rotations and controlled NOTs gates,
does not involve  measurement\cite{BBHB}.
The former approach relies on adding some ancillary quantum system in a known
state and unitarily evolving the resulting combined system.
Simply stated, a cloning machine is a device to split the information
of the input state.

When the Bu\v{z}ek and Hillery cloning machine is 
applied at the process output,
although the feedback and output of the controlled
process become imperfect, as long as the system is
controllable, so that the system can be steered to any state desired, and
observable, so that the system can be monitored,
it made no difference  for controlling.
We are no more than feeding back a transformed output.
For the feedback loop the cloner acts like an actuator
(a device added to the feedback loop to alter the system's controllability
and observability)
while for the output it distorted the output.
Although the controllability and the
observability of the 
system are modified, the coherence of the output can be preserved.
An analogy is that
having a colleague who is invariably
half-hour late for any meeting is immaterial. As long as she arrives,
it makes no difference, for all practical purposes.

\section{method}
The simplest actuator is a transformation that does nothing, which 
Wiseman called 'simple feedback'.
In practice, one can adjust the
actuator function to achieve system
controllability and observability as in classical control theories,
which is an engineering problem.

For a mixed input state,
the output state of the cloning machine reads,
\begin{equation}
{\rho}_{0,1}^{(out)}
=\frac{2}{3} {\rho}^{(in)} +
\frac{1}{6} {I},
\label{mcloner}
\end{equation}
in which $ I$ is the identity matrix. The Bloch vector shrinks by
a factor of $2/3$.

We can perform logical operations for the cloned feedback state, $B$,
with the input state, $I$, to obtain the process input
state, $M$.

In a control theory, we suppose a knowledge of only
the desired state and our input signal;
intermediate states, such as the process input, are supposed to be
unknown.
A matching condition at the process output, or any other junction
of the system, such as the process input, fixes the unknowns in
${\rho}^{(in)}$.
Controllability and observability follow if this input-output
mapping is one-to-one.

In conventional physics or engineering
practice,
a dynamic process is represented by a differential equation such
as a master equation. However,
as an equation of this kind can be written in terms of quantum operations,
we can ignore the detail and formulate the process
in turns of quantum operations\cite{NC}.

In the operator-sum formalism,
a process defined by an input density matrix
$\rho_M$, and an output density matrix $\rho_0$, with the process
described by a quantum operation, ${\cal P}$,
\begin{eqnarray}
\rho_M \stackrel{{\cal P}}{\rightarrow} \rho_0.
\end{eqnarray}
${\cal P}$ can be rewritten as a completely positive linear
transformation acting on the density matrix
\begin{equation}
{\cal P} (\rho_M)=\sum_i P_i\rho_M P_i^\dagger\;,
\label{AnklesOfHair}
\end{equation}
in which the $P_i$ satisfy the completeness relation
\begin{equation}
\sum_i P_i^\dagger P_i = I\;,
\label{HankThoreau}
\end{equation}
or equivalently  $\tr [{\cal P} (\rho_M )]=1$.

This way of treating the process is elegant.
Furthermore, we can treat the cloning machine consistently.

\section{Example}
As an example,
we consider the process of a two-level atom
coupled to the vacuum undergoing spontaneous emission.
The coherent part of the atom's evolution has a 
Hamiltonian $H= - \omega \sigma_z /2$, in which $\omega$ is the 
energy difference of the
atomic level.
The emission process is described by an Lindblad operator
$\sqrt{\gamma^{'}} \sigma_-$, in which $\sigma_- = \ket{0}\bra{1} $
is the atomic
lowering operator, and $\gamma^{'}$ is the rate of spontaneous emission.
The master equation reads,
\begin{equation}
\dot{\rho^{'}} =-i[H,\rho^{'}]+\gamma^{'}\left[
2 \sigma_- \rho^{'} \sigma_+ - \sigma_+ \sigma_- \rho^{'} - 
\rho^{'} \sigma_+ \sigma_-
\right] ,
\end{equation}
with $\sigma_+ = \sigma_-^{\dagger}$.
The solution of
this equation in the interaction picture
can be written in the operator-sum formalism, after a change of variable
$\rho(t) =\exp(i H t) \rho^{'} (t) \exp(-i H t)$ and using a Bloch
vector representation for $\rho$, as
\begin{equation}
\rho_0 = {\cal P} (\rho_M) = P_0 \rho_M P_0^{\dagger} + P_1 \rho_M P_1^{\dagger},
\end{equation}
in which 
\begin{equation}
P_0=\left(\begin{array}{cr}
1 & 0\\
0 & \sqrt{1-\gamma}\end{array}\right),\;\;\;\;
P_1=\left(\begin{array}{cr}
0 & \sqrt{\gamma}\\
0 & 0\end{array}\right),
\end{equation}
while $\gamma = 1-\exp(-2 t \gamma^{'})$ implies 
the probability of losing a photon.
For the suffixes of $\rho$, please refer to Fig.(\ref{fig1}b).

The most general input state for the process 
can be written in terms of Bloch sphere
representation as
\begin{equation}
\rho_M=\frac{1}{2}\left(I + \vec{a}\cdot\vec{\sigma}\right)=
{\frac 1 2} \left( 
\begin{array}{cr}
1+a_3 & a_1 - i a_2 \\
a_1+i a_2  & 1- a_3\end{array}
            \right)\;.
\label{bloch}
\end{equation}
Here, $\vec{a}=(a_1,a_2,a_3)$ is the Bloch vector of length unity or less,
and $\vec{\sigma}$ is the vector of Pauli matrices.
These intermediate variables are to be eliminated 
from the following calculations.

The action of the process on this density matrix
produces
\begin{equation}
{\cal P}(\rho_M)=
\frac{1}{2}\left(
I + \vec{b}\cdot\vec{\sigma}\right)\;,
\end{equation}
in which
\begin{equation}
\vec{b}= \left(a_1 \sqrt{1-\gamma}, a_2 \sqrt{1-\gamma}, 
a_3 (1-\gamma) + \gamma \right) 
\label{ab}
\end{equation}
is computed according to Eq.(\ref{AnklesOfHair}).
This state can be fed into the cloner according to Eq.(\ref{mcloner}).
The later state is also the output state of the system, since
the cloning machine we used is symmetric.

For simplicity and demonstration, we consider a system without even
the input terminal,
in the first instance.
Therefore, Bloch vector $2/3 \vec{b}$ becomes the input Bloch vector of the
process.
Hence, $\vec{a}=2/3\vec{b}$ fixes the unknown variables of the
system.
The solution is $\vec{a} =
2 \vec{b}/3 == (0,0,2 \gamma / (1+2 \gamma))$. 

If one want to have an input signal to control the system, at least
a two-qubit gate is needed for the gate, $G$, to combine the feedback signal
and the input signal. It can be a controlled-NOT gate or a controlled-phase
gate for example.
Even more generally, a controlled-U gate\cite{EHI} can serve. 
Furthermore, according to the Landauer's principle,
if one 
wishes the system to be reversible,
the gate must has input terminals and output terminals of equal number, 
as shown in the figure,
although
the extra output signal can be discarded.

Suppose we take the input signal, $I$, 
having a Bloch vector $\vec{i}=(e_1, e_2, e_3)$, 
and the feedback signal, $B$, as the control signal.
The controlled-NOT gate read,
\begin{equation}
C_{NOT}=
\left( 
\begin{array}{lccr}
0 & 1 & 0 & 0\\
1 & 0 & 0 & 0\\
0 & 0 & 1 & 0\\
0 & 0 & 0 & 1 
\end{array}
            \right)\;.
\end{equation}
The total density matrix of the input and the feedback state
is a direct product of the input density matrix, $\rho_I$, and
the feedback density matrix, $\rho_B$.
After they interact through $C_{NOT}$,
partial trace is taken over the $I$ degree of freedom. The resulting 
density matrix reads
\begin{equation}
\tr_{I} ( C_{NOT} \cdot  \rho_{I} \otimes \rho_B  \cdot C^\dagger_{NOT}) =
{\frac 1 2}
\left(
\begin{array}{lr}
1 + {\frac {2 } 3} e_3 ( a_3 (1-\gamma)+\gamma)    
& 
{\frac {2 \sqrt {1-\gamma}} 3} (a_1 -i a_2 e_3 )
\\
{\frac {2 \sqrt {1-\gamma}} 3} (a_1 +i a_2 e_3 )
& 
1 - {\frac {2 } 3} e_3 ( a_3 (1-\gamma)+\gamma)    
\end{array}
\right).
\label{match}
\end{equation}
Only one component of the input Bloch
vector has an influence on the process,
because the
$C_{NOT}$ gate mixed the total density matrix only slightly whereas the
partial trace operation eliminated all other components.

Matching Eq.(\ref{bloch}) and Eq.(\ref{match})
yields
$a_1=a_2=0$, and $a_3=e_3\gamma/(3/2-e_3(1-\gamma))$, 
which fixes the unknown variables of the process.
A valid Bloch vector has to be of length unity or less.
Furthermore, note $\gamma$ is also less than unity.
Through  Eq.(\ref{ab}),
one obtains the input-output relationship of the system:
$b_1=b_2=0$, and $b_3=\gamma/(1-(2/3)e_3(1-\gamma))$.
The system is controllable in the valid range of the process
because the mapping between
$\vec a$ and $\vec i$ is one to one.
Furthermore, according to
Eq.(\ref{ab}) the system is observable, because
we can calculate the process state once we know the output state.

Similar calculation can be done
if one take $I$ as the control signal.

In summary, the system considered maps states with
Bloch vector $\vec{i}$ into
states with 
Bloch vector $2/3\vec{b}$. If the mapping from $\vec{i}$ to $\vec{b}$
is one to one,
by definition, the system is
controllable\cite{TK}.
The mapping of state with Bloch vector
$2/3 \vec{b}$ 
to state with Bloch vector
$\vec{a}$
is one to one signifies
the system is observable.

\section{discussion}

Coherent feedback is formulated without using
adiabatic elimination as Wiseman did.
Our method indicated that,
although the system cannot be divided, the information
can. Therefore part of the output information is feed back for control
purposes.

This work is a synergy of two fields
in modern science, namely, automatic control theories and
quantum-cloning theories.
As Bru{\ss} {\it et al.} mentioned,
presently quantum-cloning theories have been
mainly of academic interests\cite{BCL}. 
A concrete example
of the application of cloning theory is presented here.

In the example given, one has the freedom to either feedback
the ancilla or one of the cloned copy, since the ancilla also contains
the process information\cite{BCL}.
According to the figure, their roles are similar.

The input control signal need not be another unknown
quantum state\cite{JAZB}. It can be a quantum system in its eigenstate.
We can change other parameters, such as time, for controlling.
The details depend on the design of the system.
On the other hand, an unknown state controlling another unknown state
is not necessarily useless: A NOT-gate is an example\cite{B}.
In the same way we can design gates that switch the quantum state
to a particular state.  
For instance, a gate that always rotates the
input state by 45 degree.
A control system which incorporated 45, 90, 135 degrees of rotation, 
with selection
is another way to achieve
quantum control.

With
a quantum feedback system formulated in this way, many recent 
results in quantum information sciences are applicable. For instance,
the system can be considered as a channel; it entropy exchange
can be calculated and the uncertainty principle derived\cite{TL}.

Our formalism does suffer from some  deficiencies:
Firstly, the outputs of the cloner are likely to entangled with
themselves or with the ancilla\cite{BM}. 
We assume this does not happen presently.
Secondly,
because the operator-sum formalism ignores 
the detail evolution of the system,
we can take no time lag between various components into account.
The  evolution time of any component must be finite\cite{GLM}.
The entanglement cannot be eliminated
in view of the second point
because entanglement serves to accelerate the evolution.
However, the time lag problem only makes difference at
the design stage.
The basic method for doing quantum feedback control is unchanged.

For the control scheme proposed by Lloyd\cite{L},
he  found quantum control with coherent feedback often
do not distinguish between sensors and actuators.
This is probably because of the models he considered are 
pathological.
In conventional control theories,
there is always an input control signal
to be mixed with the
feedback signal.
Therefore the roles of the
process and the
actuator are made asymmetric.
However, in Lloyd's examples this input signal is  missing.

Recently Simon {\it et al.} proposed doing quantum cloning
via stimulated emission\cite{SWZ,KSW}.
In view of our proposal,
this recently realized cloning machine appears to be part of
Warszawski and Wiseman's feedback control system\cite{WW} in that
they all have feedback in one mode which coupled
to another mode for output.
Perhaps their device has a cloner implicitly built-in?

As quantum-cloning machine has been realized recently\cite{LLHZGJ,MMB,
LSHB,CJFSMPJ}, and
a method for making cloning machine for cavity QED is proposed\cite{MOR},
the feedback
control method proposed herein should be verifiable.

Finally, I thank Dr. Giyuu Kido, chairman of APF8, and the organization
committee for generous support and hospitality during my visit to NIMS.

%

\end{document}